\documentstyle[epsfig, subfigure, lscape]{mn}

\title{In-Situ Particle Acceleration in Extragalactic Radio Hot 
Spots: Observations Meet Expectations}
\author[G. Brunetti et al.]
       {G. Brunetti,$^{1}$
       K.-H. Mack,$^{1,2,3}$ M.A. Prieto,$^{4}$ S. Varano$^{5,1}$\\  
       $^1$Istituto di Radioastronomia del CNR, Via Gobetti 101,
       I-40129 Bologna, Italy\\
       $^2$ASTRON, Postbus 2, NL-7990 AA Dwingeloo, The Netherlands\\
       $^3$Radioastronomisches Institut der Universit\"{a}t Bonn,
         Auf dem H\"ugel 71, D-53121 Bonn, Germany\\
       $^4$Max-Planck-Institut f\"ur Astronomie, K\"{o}nigstuhl 17,
       D-69117 Heidelberg, Germany\\
      $^5$Dipartimento di Astronomia, 
        Universit\`a di Bologna, Via Ranzani 1, I-40127 Bologna, 
       Italy\\
}

\begin{document}
\maketitle

\begin{abstract}

We discuss, in terms of particle acceleration,
the results from optical VLT observations of hot spots
associated with radio galaxies.
On the basis of observational and theoretical grounds,
it is shown that: 
1. relatively low radio-radio power hot spots are the optimum 
candidates for being detected at optical waves. This 
is supported by an unprecedented
optical detection rate of 70\% 
out of a sample of low radio power hot spots.
2. the shape of
the synchrotron spectrum of hot spots is mainly determined
by the strength of the magnetic field in the region.
In particular, the break frequency,
related to the  age of the oldest electrons in the hot spots,  
is found to increase with decreasing 
synchrotron power and magnetic field strength.
Both observational results are in agreement with 
an in-situ particle acceleration scenario.

\end{abstract}


\section{Introduction}

It is generally assumed that relativistic particles in powerful
radio galaxies and quasars, initially generated in the region 
of the active nucleus and channeled through the radio jets out to hundreds of
kpc, are re-accelerated in the radio hot spots (HSs).
These regions mark the location of strong planar shocks formed at the 
heads of the jets by their interaction with the intergalactic medium  
(e.g., Begelman et al. 1984).
There is evidence for electron acceleration in these regions, including 
their spectral energy distribution (SED), morphology and polarisation
(e.g., Meisenheimer 2003, for a review).
One of the most important indications in favour of electron acceleration 
in the HSs is still the detection of
synchrotron emission at optical bands from
high energy electrons (Lorentz factor $\gamma > 10^5$) with
very short radiative lifetimes (e.g., Meisenheimer et al. 1997).
On the other hand, it has been
argued that the optical detection of radio HSs 
still does not make in-situ acceleration an inescapable
mechanism. This is possible if one assumes 
a minimum loss scenario in which 
the relativistic electrons, injected in the nuclear region,  
flow in the jets losing energy only
via inverse Compton (IC) 
scattering of Cosmic Microwave Background
(CMB) photons
(Gopal-Krishna et al. 2001).
However, the detection of the optical
emission in the southern HS of 3C\,445
{\it resolved} in a complex extended structure
(Prieto, Brunetti \& Mack 2002) 
and the large-scale diffuse optical emission
in Pictor A-West (Meisenheimer 2003) 
are difficult to be accounted for by a 
simple minimum loss transport scenario.

\noindent
The purpose of this Letter is to test some specific 
predictions of the in-situ acceleration scenario
with direct observations of a relatively large number
of HS regions for which reliable SEDs 
could be determined.
In brief, we discuss two predictions of the theory:
1. HSs of relatively low radio power, i.e. most likely
dominated by low magnetic field values,
are the {\it optimum} candidates for being detected at optical
wavelengths;
2. the shape of the HS 
synchrotron spectra depends on the value
of the magnetic field in these regions.

\noindent
To that end, we have observed 
in the optical a sample of 
HSs with radio power lower 
by about one order of magnitude with respect to 
that of typical
known HSs.
This selection criterion, derived from simple
arguments based on particle acceleration theory, 
led to a 70\% detection rate in the optical
with the VLT (Mack et al. 2003).
In addition to our VLT sample, in this Letter we have 
compiled a data set of SEDs, 
from radio to optical, of a few optical synchrotron
HSs from the literature and compared their
properties with that of our VLT HSs.

\noindent
$H_0=50$ km s$^{-1}$ Mpc$^{-1}$ and $q_0=0.5$ are used in
the paper.

\section{Electron transport}

Relativistic electrons in powerful radio sources
are produced in the nuclear regions.
Under realistic assumptions for the spatial
transport, the diffusion velocity
of the fast particles in the jet is expected to 
be considerably lower than the advection velocity 
(e.g., Jones et al. 1999; Casse \& Marcowith 2003)
and thus these particles are expected to be 
channeled and advected into the jet 
up to the HSs.
HSs are often located at several tens
or hundreds of kpc from the nucleus of the radio
galaxy  
and thus the travel time of the electrons up to the HS
can be considerably
larger than the radiative lifetime of these
electrons, in particular when they are detected at optical frequencies.

\noindent
One possibility is 
that the radiative losses of the electrons in the jet
are minimized.
This assumes that adiabatic and synchrotron losses
in the jets are negligible with respect to the unavoidable
losses due to IC scattering of the CMB photons.
In the framework of 
this minimum loss scenario, Gopal-Krishna et al.~(2001) 
have indeed shown that, at least
in some cases, it is possible
to transport high energy electrons in the jet before
they have lost a substantial fraction of their
energy.
More specifically, if electrons in the jet are transported with a 
bulk velocity, $c \beta_{j}$, the projected 
maximum distance from the nucleus, $D_{max}$, 
at which it is still possible to produce 
optical emission is:

\begin{equation}
D_{max} (kpc) \simeq
43 \times 
{{ \Gamma_j \beta_j \sin \psi (1+z)^{1/2} }\over{
0.09 (1+z)^4 \Gamma_j^2 +
( {{ B_j}\over{10}} )^2
}}
\Big(
{{B_{HS} }
\over {100}}
{{ 10^{15}}\over
{\nu_o}}
\Big)^{1/2}
\label{maxd}
\end{equation}

\noindent
where $\Gamma_j$ is the Lorentz factor of the jet, 
$B_{HS}$ and $B_{j}$ are the HS and jet 
magnetic fields (in $\mu G$) respectively, 
$\nu_o$ is the observed synchrotron frequency (in Hz), 
and $\psi$ is the angle between jet velocity 
and line of sight.

\noindent
In Fig.~1 we show the maximum accessible distances for the electrons 
for different values of their bulk velocity in the jet and  
under a reasonable choice of values for the other parameters in Eq.~1 
(see caption Fig.~1).
Fig.~1 shows that the maximum possible distance, of the
order of 150 kpc (for $B_{HS}=100 \mu$G), is achieved for the
extreme case of $B_{j}=0$.
We notice that the presence of a 
low magnetic field in the jet, $B_j \sim 5 \mu$G, considerably
reduces $D_{\rm max}$.

\noindent
As a general result, Eq.~1 yields 
$\nu_o \propto B_{HS}/D_{max}^2$, i.e. 
the detection of high-frequency synchrotron emission is
expected in HSs at small distance from the nucleus and
with high magnetic field strengths.

\begin{figure}
\resizebox{\hsize}{!}{
\includegraphics{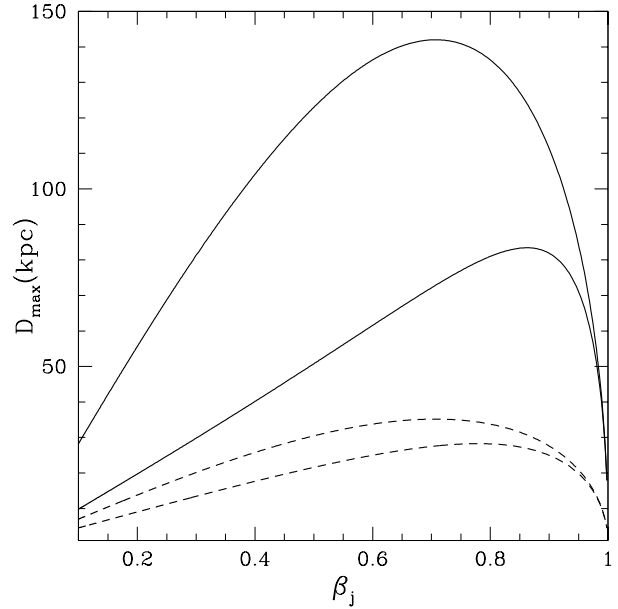}}
\caption[]{
The maximum projected distance electrons can travel
from the nucleus (Eq.~\ref{maxd}) 
still emitting synchrotron radiation
at $\nu_o = 10^{15}$ Hz is reported for
z=0.1 (solid lines) and z=0.5 (dashed lines).
$B_j=5\;\mu$G (lower curves)
and $B_j=0$ (upper curves) are
assumed, with $\psi=1.15$ and $B_{HS}=100\;\mu$G.
}
\end{figure}

\section{Particle acceleration}

If optical HSs are detected beyond the maximum distance achievable with 
a transport scenario, the high energy electrons responsible for that 
emission must have been accelerated in-situ, most likely via shock 
acceleration (e.g., Blandford \& Eichler 1987).
The shape of the electron 
spectrum accelerated in a shock region basically 
depends on the interplay of the acceleration efficiency
with the energy losses of the electrons and with the rate of the
electron diffusion from the shock region.
In case of non-relativistic, diffusive shocks, 
the spectrum of the accelerated electrons accumulated in the
HS volume is well known: a power law in 
momentum, $N(p) \propto p^{-\delta}$, up to 
a break, $p_b = m c \gamma_b$, then a steepening 
up to maximum momentum, $p_c = m c \gamma_c$,
where a cut-off is developed  
(e.g., Heavens \& Meisenheimer 1987).
The cut-off energy is generated by the competition between
acceleration and loss mechanisms in the shock region, while
the break energy is the maximum energy of the oldest electron
population in the HS volume and its value is driven by the cooling
of the electrons in the post-shock region.

\noindent
Particles are accelerated in the shock region 
crossing back and forth between upstream
and downstream flows.
The scattering should be guaranteed by strong turbulence
in the jet flows which 
can be externally produced or 
excited by the particles themselves (e.g., Cesarsky 1980).
The efficiency of the acceleration depends on the spectrum of
the turbulent waves in the flows
due to the spatial diffusion coefficient
of particles ${\cal K}(\gamma)$ (e.g., Casse \& Marcowith 2003
and ref. therein).
The rate of electron energy gain is given by:

\begin{equation}
\left( {{ d \gamma }\over{d t}}\right)_{\rm sh}^+
\simeq \gamma {{ {U_-}^2 }\over{r}} \left( {{ r-1}\over{ r+1}}
\right) {1 \over {3 {\cal K}(\gamma) }}
\label{shock}
\end{equation}

\noindent
where 
$U_-$ is the velocity of the plasma in the downstream region 
and $r$ is the compression factor.
For sake of simplicity, most theoretical applications to HSs done so far
assume a constant diffusion coefficient (e.g., Heavens \& Meisenheimer,
1987).
However, because of its dependence on the 
topology of the turbulent magnetic field in the shock region 
(e.g., Blasi 2001 and ref. therein), 
here, we focus on two relevant diffusion coefficients:
a Kolmogorov coefficient, obtained assuming a Kolmogorov
spectrum for the magnetic field in the upstream and downstream
region, and a classical Bohm diffusion coefficient
(e.g., Bhattacharjee \& Sigl 2000, for a review):

\begin{equation}
{\cal K}(\gamma)
\simeq 
\left\lbrace \begin{array}{lll}
1.3 \times 10^{29} (B_{\mu G})^{-1/3}
L_{\rm kpc}^{2/3}
\gamma^{1/3} & \textrm{Kolmogorov} \\
      &       \\
1.7 \times 10^{19} (B_{\mu G})^{-1} \gamma
& \textrm{Bohm}
\end{array}
\right.
\label{spatial_d}
\end{equation}

\noindent
with $L_{kpc}$ the maximum coherence scale of the magnetic
field in kpc (of the order of the HS size).

\noindent
On the other hand, 
the rate of synchrotron and IC losses in the shock 
region is given by :

\begin{equation}
\left( {{ d \gamma }\over{d t}}\right)_{\rm syn+ic}^-=-
{\cal S} \gamma^2 
\Big( {2\over 3} B^2 
+ B_{IC}^2
\Big)
\label{syn+ic}
\end{equation}

\noindent
where ${\cal S} \simeq 1.9 \times 10^{-9}$, and 
$B$ and $B_{IC}$ are the magnetic
and IC equivalent field (in this case 
in the shock region).
At the zeroth order, i.e. without including non-linear
effects,
the maximum energy of the accelerated electrons,
$\gamma_c$, is the energy at which cooling is
balanced by acceleration, i.e. (from
Eqs.(\ref{shock}-\ref{syn+ic})):

\begin{equation}
\gamma_{\rm c}
\sim
\left\lbrace \begin{array}{lll}
1.2 \times 10^5 
{{
f^{3/4}
}\over{
L_{\rm kpc}^{1/2}
}}
\Big(
{{ U_-}\over{0.1 c}}
\Big)^{3/2}
\Big[
{{ ( B/{(100 \mu G)} )^{1/3} }\over
{ 2/3 (B/{(100 \mu G)})^2+
(B_{\rm IC}/{(100 \mu G)})^2
}} \Big]^{3 \over 4}    \\
                 \\
10^9 \sqrt{f}
\Big(
{{ U_-}\over{0.1 c}}
\Big)
\Big[
{{ B/{(100 \mu G)} }\over
{ 2/3 (B/{(100 \mu G)})^2+
(B_{\rm IC}/{(100 \mu G)})^2
}}
\Big]^{1 \over 2}
\end{array}
\right.
\label{max-kolm}
\end{equation}

\noindent
in the Kolmogorov and Bohm case, respectively
($f=(r-1)/(r^2+r)$, $r=4$ for a strong shock).
Eq.~\ref{max-kolm} shows that 
for fiducial values of the parameters
electrons with $\gamma \geq 10^5-10^6$ 
are easily obtained, and therefore optical emission is guaranteed
regardless of the distance from the core.
We also notice that the corresponding cut-off 
frequency, $\nu_c \propto \gamma_c^2 B$,
would increase with
decreasing magnetic field strength (Kolmogorov case) or
would be roughly constant (Bohm case). 

\noindent
Once the electrons are accelerated, they travel 
in the post-shock region filling the HS volume
and cooling due to radiative losses
but also due to adiabatic expansion.
Without a detailed model for the HS dynamics, 
the break energy of the electrons can be simply
parameterized as: 

\begin{equation}
\gamma_b (\tau) = 
{{
\gamma_c }\over{
1 + {\cal S} \langle B^2_{\tau} \rangle \gamma_c \tau 
}}
\times
\Big(
{{ u_{l} }\over{ u_{HS} }}
\Big)^{1/4} 
\label{break}
\end{equation}

\noindent
where $\tau$ is the dynamical age of the HS,
$\langle B^2_{\tau}\rangle$ gives
the synchrotron and IC fields
averaged over the electron time evolution,
and $(u_{l}/u_{HS})^{1/4}$ accounts for
adiabatic losses (e.g.,
Manolakou \& Kirk 2002; $u_l$ and $u_{HS}$ are the
energy density in the lobe and HS, respectively) 
under the rough assumption
that, at a given point, the HS expands into the lobe in a time
shorter than the radiative cooling time.
The corresponding synchrotron break
frequency emitted in an averaged
HS field, $B_{HS}$, is thus 
(using Eq.~\ref{break}):

\begin{equation}
\nu_b
\propto
{{ \gamma_c^2
B_{HS} (u_{l}/u_{HS})^{1/2} 
}\over{
\left(
1+ {\cal S} \tau \langle B^2_{\tau} \rangle
\right)^2
}}
\rightarrow
{{(u_{l}/u_{HS})^{1/2}}\over{ \tau
^2}}
B_{HS}^{-3} 
\label{nub}
\end{equation}

\noindent
for synchrotron dominated ($B \gg B_{IC}$) HSs.
The right side of Eq.~\ref{nub}, 
(valid for $\gamma_c \gg \gamma_b$) is obtained
for $\langle B_{\tau} \rangle \sim B_{HS}$.

\noindent
Thus, Eqs.~\ref{max-kolm}-\ref{nub} make clear predictions on the
shape of the spectrum of the
shock-accelerated electrons in HSs.
In particular, 
{\it the weaker the magnetic field, the higher are the 
break and the cut-off frequencies}.
As a consequence, low-field HSs are 
the most promising candidates for being detected at
optical wavelengths.

\begin{figure}
\resizebox{\hsize}{!}{
\includegraphics{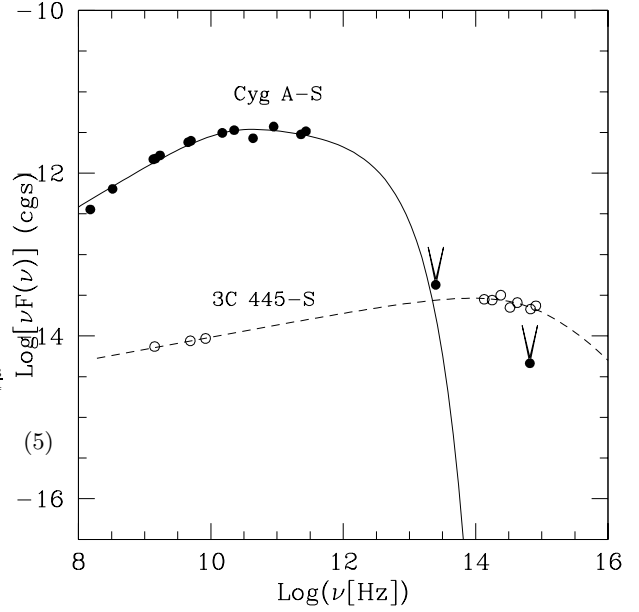}}
\caption[]{
The spectrum of the southern HS of Cyg A 
(filled points and arrows - upper limits) is compared
with that of the low-brightness
southern HS of 3C\,445 
(empty points).
The curves are theoretical 
synchrotron models.
The data of Cyg\,A South are taken from Carilli
et al.~(1999).
}
\end{figure}

\section{VLT Sample and results}

\subsection{Optical Detection rate}

The search for optical counterparts of HSs has so far been addressed 
to bright radio HSs. 
However, only a few optical HSs were discovered in more than 20 years 
of observations. 
Because of this scarcity of detections, the exploration of low brightness
HSs has never been attempted.
However, as shown in Sect.~3, once the electrons
are injected/accelerated in the
HS volume, their radiative life-time, in low-power
radio HSs (most likely with lower values of magnetic
field strength), 
is expected to be longer than that in high-power
HSs. Thus, low-power radio HSs would have higher $\nu_b$ and 
emit the bulk of their radiation 
at high frequencies.
This is sketched in Fig.~2 where 
a comparison between the spectrum of 
a high and low-power HS is given: if the general expectations derived 
in Sect.~3 hold,  
the spectra of low-power HSs are 
expected to remain relatively hard at high frequencies 
and thus, contrary to the case of
the high-power ones, 
the optical detection rate of 
low-power targets is expected to be relatively high.

\noindent
Thus, to test these expectations and 
to extend the search for optical counterparts down to relatively 
low-power HSs, we embarked 
on a project with the VLT aimed at the detection of HSs
with synchrotron powers about one order of
magnitude lower 
than those of typical optical HSs reported in the literature.
Ten HS targets were selected from Tadhunter et al. (1993). 
Although relatively faint in the radio band, 
these targets turned out
to be sufficiently bright in the optical 
to warrant a significant detection with the VLT 
under the assumption that their break frequencies fall in 
between the far-IR and optical band (assuming a
synchrotron spectral index $\alpha \sim 0.6-0.7$;
$P(\nu)\propto \nu^{-\alpha}$).
Out of the 10 HSs observed with the VLT, 
clear detections were achieved 
for six of them (3C\,105 South, 3C\,195 South, 
3C\,227 West, 3C\,227 East, 3C\,445 North and 3C\,445 South), 
while one of the HSs in the sample, 3C\,327 East, 
is confused by a relatively bright spiral galaxy which hampers 
the detection of any HS counterpart.
All together, an extremely high detection rate of 70\% was 
achieved which basically confirms, at least qualitatively,
our theoretical expectations.

\begin{figure}
\resizebox{\hsize}{!}{
\includegraphics{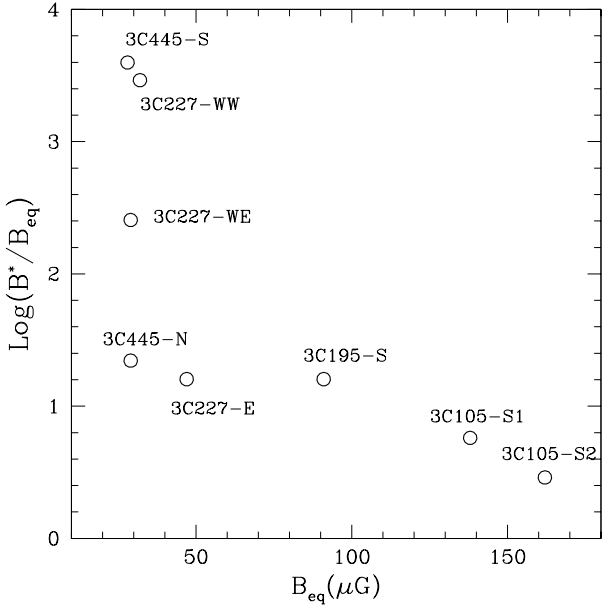}}
\caption[]{
The ratio $B^*/B_{eq}$ as a function
of $B_{eq}$ calculated 
for the eight HS components detected with
the VLT.
Note that the new R, B and U data for 3C\,445-S 
led to an improved determination of the SED with respect 
to the results of Prieto et al.~(2002):
$\nu_c \geq 10^{16}$ Hz is obtained
(here $\nu_c=10^{16}$ is assumed).
}
\end{figure}

\subsection{Spectral Parameters and Particle Acceleration}

In order to derive 
the spectral parameters (i.e., $\alpha= (\delta-1)/2$, 
the break, $\nu_b$, and the cut-off, $\nu_c$, frequencies) of 
the HSs, theoretical synchrotron spectra produced by 
a population of accelerated electrons (following the modelling 
in Brunetti et al. 2002) were fit to the data
(possible effects due to beaming (Georganopoulos \&
Kazanas 2003) are not included for simplicity).
To this aim we have collected broad-band data for the HSs of our sample.
In particular, VLA high-resolution images were 
obtained either from the NRAO-VLA archive or from new observations
by us. 
In case of two HSs (3C\,105 South and 3C\,227 West) two well separated
optical components, well matching the corresponding 
separated radio counterparts (Mack et al. 2003), were detected. 
Thus, the final number of VLT counterparts was equal to eight.

\noindent
In the framework of the minimum-loss transport scenario, 
assuming a HS at a projected distance from the nucleus, $D_{HS}$,
the HS magnetic field, $B^*$, required to obtain synchrotron
emission up to the cut-off frequency $\nu_c$ is obtained
from Eq.~1 :

\begin{equation}
{{ B^*(\nu_c) }\over{
100 \mu G}}
\simeq
\Big(
{{\nu_c}\over { 10^{15}}}
\Big)
\Big(
{{ D_{HS} }\over{43 kpc}}
\Big)^2
\Big[
{{0.09 (1+z)^4 \Gamma_j^2 +
( {{ B_j}\over{10}} )^2
}\over{
\Gamma_j \beta_j \sin \psi (1+z)^{1/2}
}}
\Big]^2
\label{b*}
\end{equation}

\noindent
Assuming a conservative value of the jet magnetic field 
$B_{j}\sim 5\;\mu$G, 
in Fig.~3 we report the ratio $B^*/B_{eq}$
($B_{eq}$ the equipartition field, computed
using formulae with a low energy
cut-off, e.g. Brunetti et al. 1997)
calculated for all the HS components detected in our VLT sample.
It can be seen that
in the case of HSs with lower $B_{eq}$, 
extremely large fields 
($B^* \sim 10-10^4 B_{eq}$) 
should be admitted at least in a considerable fraction 
of the emitting volume 
of the HSs to produce the observed optical emission.
This is clearly because, under realistic physical conditions, 
the travel time from the nucleus to the HS region
of the high-energy electrons emitting
in the optical band is 
considerably longer than their 
cooling time.
As a consequence, the results in Fig.~3 
indicate the presence of in-situ particle
acceleration at least for the most extreme cases.

\subsection{B-field and break frequency}

The high optical detection rate derived in Sect.~4.1
for low-power HSs indicates that the break frequency
of these HSs is larger than that of the high-power
counterparts for which only a very low detection 
rate has been achieved in the last years.
From Eq.~\ref{nub}, this 
could be due to either the effect of stronger $B$ in powerful
HSs (Blundell et al. 1999)
or possible smaller emitting volumes
(i.e., smaller $\tau$ in Eq.~\ref{nub} for a constant velocity
in the post shock region) in low-power
HSs (Meisenheimer et al. 1989).
We follow the simplest approach that low-power HSs
have a lower magnetic field strength.
Accordingly, in Fig.~4 we display the break frequency as a function of the
HS magnetic field (with $B_{HS} \sim B_{eq}$). 
Besides the fact that we deal with only 8 detections and 3
upper limits, a relatively clear trend between
$\nu_b$ and $B$ is found, suggesting that 
the large values of the
break frequency found in the case of low-power
HSs is most likely due to the effect of $B$ on the
cooling of the electrons.
Indeed, assuming a roughly constant emitting volume
(or $\tau$ for a const. velocity in the post shock region), 
Fig.~4 also shows the zeroth-order 
approximation (no adiabatic losses)
of Eq.~\ref{nub} ($\nu_b\propto B^{-3}$, solid line). 
The slope of this approximation is 
surprisingly close to that observed.
With the aim to compare the behaviour of 
our VLT sample with other HSs selected to match
the same range of $\nu_b$, we extracted all the flux densities of
optical synchrotron HSs from the literature.
In Fig.~4 we display the break frequency and the equipartition
magnetic field strength derived with the same procedure 
described
below for these 8 additional optical synchrotron HSs 
(Tab.~1).
The radio and optical data from the literature 
have been combined with archive HST data when available. 
In this Letter we do not include the case of 3C\,263E
since it has been shown that the broad band 
spectrum of this HS might result from a combination of 
two different spectral components (Hardcastle et al. 2002), 
and the cases of 3C\,196N and
3C\,295N whose optical emission is best interpreted as
due to synchro-self-Compton (Hardcastle 2001; Brunetti 2000).
We notice that
the behaviour followed by the HSs from the
literature is consistent with that of the VLT ones and thus
strengthen the derived $\nu_b - B_{eq}$ trend.
\begin{table}
\begin{center}
\caption{Optical HSs from the literature}
\begin{tabular}{ccccc}
\hline
Name & z    & $\log({\rm P}_{bol}$) & $\log(\nu_b$)$^{a}$ & Ref$^{b}$ \\
     &      &   [erg/s]        &     [Hz]     &     \\
\hline
      3C\,20W &0.174  &42.92    & 13.0     & [1][2]\\
      3C\,33S &0.0592 &42.13    & 12.65    & [1][2]\\
      3C\,111E&0.0485 &42.04    & 13.26    & [1][2]\\
      3C\,303W&0.141  &42.4     & 13.9     & [1][2]\\
      3C\,351L&0.372  &43.5     & 12.5-13.3& [3][4]\\
      3C\,351J&0.372  &43.4     & 12.1-12.4& [3][4]\\
    3C\,390.3N&0.056  &41.5     & 13.2     & [5][6]\\
      Pic\,A-W&0.0361 &42.36    & 14.0     & [1][2]\\
\hline
\end{tabular}
\end{center}

$^{a}$ 
With the exception of 3C\,351 errors on the break frequency
are within a factor of 2.
$^{b}$
[1] Meisenheimer et al., 1989; [2] Meisenheimer et al., 1997;
[3] Brunetti et al., 2001; [4] Hardcastle et al., 2002;
[5] Prieto \& Kotilainen, 1997; [6] Harris et al., 1998.
\end{table}
It is clear that such a trend should be tested combining the
VLT sample with a sample of HSs with relatively high values
of the magnetic field and thus with an
{\it expected} synchrotron
break frequencies below $\sim 10^{12}$ Hz (Fig.~4).

\begin{figure}
\resizebox{\hsize}{!}{
\includegraphics{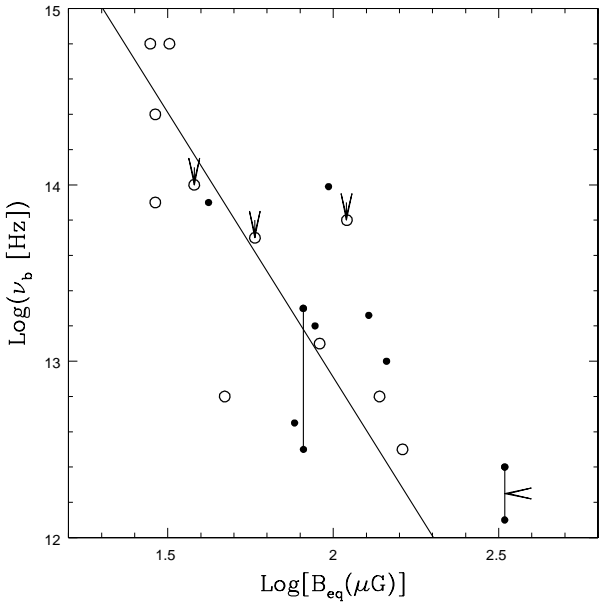}}
\caption[]{
The synchrotron break frequency as a function of the
equipartition magnetic field for the VLT HSs (empty points),
and for comparison, for optical HSs from the literature (filled
points). 
The solid line model is $\nu_b \propto B^{-3}$ (see text).
The upper limit in $B_{eq}$ is for 3C\,351J 
for which there is some indication that the real  
magnetic field is lower than the equipartition one 
(Brunetti et al., 2001; Hardcastle et al., 2002).
The large uncertainties for the $\nu_b$ of
3C\,351L and 351J are caused by the errors in the 
HST R-band fluxes.
}
\end{figure}

\section{Conclusions}

VLT observations of a sample of relatively low-radio 
power HSs has led to an unprecedented 70\% detection rate
of these regions in the optical.
From a theoretical point of view, this result is the natural 
consequence of optical emitting electrons surviving longer in 
low power, i.e. low-$B$ field, 
regions than in bright and strong-field HSs.

We have shown that the combination of low magnetic fields and
large distances from the core of the HSs in our VLT
sample indicates the need for in-situ particle
acceleration in these regions: under physically realible
hypotheses, the emitting electrons are
too energetic and distant from the core
to be transported from the nucleus to the 
HS region.

We have shown that there is a trend between the magnetic field
strength (equipartition) in the HS volume and the measured
synchrotron break frequency for the HSs in our VLT sample.
In addition, we also have shown that the behaviour
of optical synchrotron HSs taken from the literature is
consistent with that of our VLT HSs.
This trend has the {\it predicted shape} inferred within 
the framework of 
the in-situ acceleration scenario in which the break energy
of the accelerated/injected 
electrons is driven by the 
cooling in the post-shock region.
The combination of our low power sample
with statistical samples of high power HSs
will test the reported trend: the break frequency
of high power HSs is in general expected to be 
below $\sim 10^{12}$ Hz.

\section{Acknowledgements}

We thank K. Meisenheimer, J. Kirk and
G. Setti for useful discussions and the anonymous 
referee for helpful advice on the presentation of the
paper.
GB and KHM acknowledge the MPIA for their warm hospitality
and partial financial support.
GB and SV acknowledge partial financial support from
MIUR under grant COFIN 2001-02-8773.
KHM was supported by a Marie-Curie-Fellowship of the European
Commission.
This paper
is based on observations made with the VLT and on data 
obtained from the HST data 
archive.


\begin{thebibliography}{}
\bibitem{} Bhattcharjee P., Sigl G., 2000, Phys. Rep. 327, 109
\bibitem{} Begelman M.C., Blandford R.D., 
 Rees M.J., 1984, Rev. of Mod. Phys. 56, 255
\bibitem{} Blandford R., Eichler D., 1987, Phys. Review B
 154, 1
\bibitem{} Blasi P., 2001, APh 15, 223
\bibitem{} Blundell K.M., Rawlings S., Willott C.J., 1999, AJ 117, 677
\bibitem{} Brunetti G., 2000, 
in {\it Particle and Fields in Radio Galaxies},
A.S.P. Conf. Ser., 250, 238, R.A.Laing \& K.M. Blundell Eds.  
\bibitem{} Brunetti G., Setti G., Comastri A., 1997,
 A\&A  325, 898
\bibitem{} Brunetti G., Bondi M., Comastri A., 
 Pedani M., Varano S., Setti G., Hardcastle M.J., 2001,
 ApJ 561, L157
\bibitem{} Brunetti G., Bondi M., Comastri A.,
 Setti G., 2002, A\&A 381, 795
\bibitem{} Carilli C.L., Kurk J.D., van der Werf P.P., 
 Perley R.A., Miley G.K., 1999, AJ 118, 2581
\bibitem{} Casse F., Marcowith A., 2003, A\&A 404, 405
\bibitem{} Cesarsky C., 1980, Ann. Rev. Astron. Astrophys.
18, 289
\bibitem{} Georganopoulos M., Kazanas D., 2003, ApJ 589, L9
\bibitem{} Gopal-Krishna, Subramanian P., 
 Wiita P.J., Becker P.A., 2001, A\&A 377, 827
\bibitem{} Hardcastle M.J., 2001, A\&A 373, 881
\bibitem{} Hardcastle M.J., Birkinshaw M., Cameron R.A.,
 Harris D.E., Looney L.W., Worrall D.M., 2002, ApJ 581, 948
\bibitem{} Harris D.E., Leighly K.M., Leahy J.P., 1998,
ApJ 499, L149
\bibitem{} Heavens A.F., Meisenheimer K., 1987, MNRAS 225, 335
\bibitem{} Jones T.W., Ryu D., Engel A., 1999, ApJ 512, 105
\bibitem{} Mack K.-H., Prieto M.A., Brunetti, G., 
2003, New Astronomy Rev., in press. 
\bibitem{} Manolakou K., Kirk J.G., 2002, A\&A 391, 127
\bibitem{} Meisenheimer K., 2003, New Astronomy Rev., in press. 
\bibitem{} Meisenheimer K., R\"oser H.-J., Hiltner P.R., 
 Yates M.G., Longair M.S., Chini R., Perley R.A., 1989, A\&A 219,
 63
\bibitem{} Meisenheimer K., Yates M.G., R\"oser H.-J., 1997, 
 A\&A 325, 57
\bibitem{} Prieto M.A., Kotilainen J.K., 1997, ApJ 491, L77 
\bibitem{} Prieto M.A., Brunetti G., Mack K.-H., 2002, 
 Science 298, 193
\bibitem{} Tadhunter C.N., Morganti R., di Serego-Alighieri S., 
 Fosbury R.A.E., Danziger I.J., 1993, MNRAS 263, 999
\end{thebibliography}
\end{document}